\documentclass[11pt,preprint]{aastex}

\slugcomment{ApJ in press}

\shorttitle{SMA Observations of FN Tau}
\shortauthors{Momose et al.}

\begin{document}

\title{High Resolution Observations of Dust Continuum Emission at 
340 GHz from the Low-mass T Tauri Star FN Tauri}



\author{Munetake Momose}
\affil{College of Science, Ibaraki University, Bunkyo 2-1-1, Mito, Ibaraki 310-8512, Japan.}
\email{momose@mx.ibaraki.ac.jp}

\author{Nagayoshi Ohashi}
\affil{Academia Sinica Institute of Astronomy and Astrophysics, P.O. Box 23-141, Taipei 10617, 
Taiwan, R. O. C. }

\author{Tomoyuki Kudo and Motohide Tamura}
\affil{National Astronomical Observatory of Japan, Osawa 2-21-1, Mitaka, Tokyo 181-8588, Japan.}

\and

\author{Yoshimi Kitamura}
\affil{Institute of Space and Astronautical Science, Japan Aerospace Exploration 
Agency, 3-1-1 Yoshinodai, Sagamihara, Kanagawa 229-8510, Japan. }

\begin{abstract}
FN Tau is a rare example of very low-mass T Tauri stars that exhibits a 
spatially resolved nebulosity in near-infrared scattering light. 
To directly derive the parameters of a circumstellar disk around FN Tau, 
observations of dust continuum emission at 340 GHz are carried out with 
the Submillimeter Array (SMA). 
A point-like dust continuum emission was detected with 
a synthesized beam of $\sim 0.7\arcsec$ in FWHM. From the analysis of the visibility 
plot, the radius of the emission is estimated to be $\leq 0.29\arcsec$, corresponding to 
41 AU. This is much smaller than the radius of the nebulosity, $1.85\arcsec$ for its brighter part 
at 1.6 \micron. The 340 GHz continuum emission observed with the SMA 
and the photometric data at $\lambda \leq 70 ~\micron$ are explained by 
a power-law disk model whose outer radius and mass are 41 AU and
$(0.24 - 5.9) \times 10^{-3}~M_{\sun}$, respectively, if the exponent of dust mass opacity 
($\beta$) is assumed to be $0-2$. The disk model cannot fully reproduce the flux density at 
230 GHz obtained with the IRAM 30-meter telescope, suggesting that
there is another extended ``halo'' component that is missed in the SMA observations. 
By requiring the halo not to be detected with the SMA, the lower limit to the size of 
the halo is evaluated to be between 174 AU and 574 AU, depending on the assumed 
$\beta$ value. This size is comparable to the near-infrared nebulosity, implying that 
the halo unseen with the SMA corresponds to the origin of the near-infrared nebulosity. 
The halo can contain mass 
comparable to or  at most 8 times greater than 
that of the inner power-law disk, but its surface density should be lower than that at the outer edge of the 
power-law disk by more than one order of magnitude. The physical nature of the halo is unclear, 
but it may be the periphery of a flared circumstellar disk that is not described well in 
terms of a power-law disk model, or a remnant of a protostellar envelope having flattened 
structure. 
\end{abstract}


\section{Introduction}

Hundreds of exoplanets were discovered in the last 15 years \citep{marcy05, mayor05}. 
Many planetary systems have a configuration completely different from the solar 
system, requiring us a comprehensive theory for planet formation 
\citep[e.g.,][]{ida04}. Diversity of the exoplanets may stem from a variety of the 
properties of their precursors, i.e., protoplanetary disks. \citet{kokubo02}, for 
example, proposed the idea that the planet configuration formed in a disk depends 
on its initial material distribution. Dust continuum emission at millimeter and 
submillimeter wavelengths is an important tracer to  the total mass or surface density 
of the disks. 
Indeed, a sensitive multiwavelength submillimeter continuum survey of 
153 young stellar objects in the Taurus-Auriga regions by  \citet{and05} 
revealed that circumstellar disk masses are lognormally distributed with a mean total mass of 
$5\times 10^{-3} ~M_{\sun}$ and a large dispersion (0.5 dex) when 
the gas-to-dust mass ratio of 100 is adopted. 
Meanwhile, imaging surveys of Sun-like single T Tauri stars were carried 
out with millimeter and submillimeter interferometers that can resolve disk emission 
\citep{kita02, and07}. They obtained not only typical values or evolutionary trends of 
various disk parameters but also found scatters around them, which can be interpreted as 
a diversity. 

Typical values of disk parameters or what kind of planet is likely to form must also 
depend on the stellar mass. Planet formation around stars with a mass lower than 
Sun-like stars is especially intriguing, since these are the most abundant stars in our galaxy. 
\citet{sch06} conducted sensitive 1.3 mm observations of 20 young brown dwarfs in the 
Taurus star-forming region and found that their disk masses range from 
$\la 0.4$ to several Jupiter masses, indicating that the disk mass relative to the star is 
comparable to those derived for coeval low-mass ($\leq 3~M_{\sun}$) stars, namely,  
$\la 1$~\% to 5~\% in most cases. 
They also found that at least 25\% of the targets are likely to have disks with 
radii greater than 10 AU, implying that the dynamical ejection of embryos \citep{bat02} 
might not be the dominant formation process for brown dwarfs \citep[see also][]{luh07}. 
These results may suggest that 
there may be no essential difference in the process for the formation of a
star-disk system between Sun-like stars and less massive stars including brown dwarfs. 
Compared to the Sun-like stars, however, 
the typical mass and surface density of the disks around M dwarfs and brown dwarfs are expected to be 
lower while the snow line is located at a smaller radius. Such differences in the initial condition may 
suppress the formation of gas giants but may result in higher frequency of icy 
Neptune-mass planets \citep{lau04, ida05}. Although these theoretical predictions seem to 
be supported by recent planet searches around M dwarfs \citep{bea06, joh07}, 
physical properties of the disks around their younger counterparts should be 
studied further to verify the above considerations.  

Another important clue for understanding the planet formation process in the disks 
is dust grain property such as composition, crystallinity, and size distribution. These have been studied 
by spectroscopic observations of silicate emission features at mid-infrared wavelengths 
and have been observed toward many young stars ranging from brown dwarfs to 
Herbig Ae/Be stars \citep[for review, see][]{nat07}.  
For Sun-like stars, for example, \citet{sar06} analyzed $8-14~\mu$m emission 
spectra of 12 T Tauri stars in the Taurus-Auriga dark clouds and TW Hydrae obtained with 
Infrared Spectrograph (IRS) on board the Spitzer Space Telescope. They found that 
later spectral type stars can have relatively large amounts of 
crystalline silicates in their surrounding disks. More recently, a
comparative study of the dust and gas properties of disks around young 
Sun-like stars (K1-M5) and cool stars including brown dwarfs (M5-M9) was 
made by \citet{pas09}. They revealed that lower-mass stars including brown dwarfs 
tend to show weaker features at $10 ~\mu$m, indicating that the features are dominated 
by more ``processed'' dust grains than those from disks around 
higher-mass Herbig Ae/Be stars \citep[see also][]{apai05}. 
These systematic trends may be related to the fact that the location of the 
10~$\mu$m silicate emission zone is closer to the star as the stellar luminosity is 
lower; the larger crystallinity seen toward cooler stars can be accounted for if the size of 
dust grains in the inner regions is systematically larger  
because the contribution to the spectrum from amorphous silicates diminishes 
more rapidly with grain growth ($0.1-10~\mu$m) than that from crystalline 
silicates do \citep{kes07}. 
Although these features arise from the optically thin surface layer of the inner regions 
of the disks and might not necessarily reflect the properties of the bulk silicate dust 
in the observed disk, these are the direct evidence for dust processing (i.e., dust growth 
and crystallization), which can be regarded as the first step to the formation of planets.  

\object{FN Tau} (\object{Haro 6-2}) is a relatively rare example of low-mass T Tauri stars whose 
near-infrared scattering light shows clear appearance of disk-like morphology \citep{kudo08}. 
It is a single star with a spectral type of M5, located at the distance of $\sim$ 140 pc. 
The stellar mass and age are estimated to be 0.11 $M_{\sun}$ and $< 10^5$ yr, respectively 
\citep{muz03}. Existence of a circumstellar disk was suggested by excess emission at infrared 
and millimeter wavelengths \citep{strom89, bec90}. Significant fraction of crystalline silicate grains
was inferred from mid-infrared spectroscopic observations, as commonly be seen in the cases of other cool 
T Tauri stars \citep{sar06, for04}, suggesting that grain processing is taking place at least in the disk 
inner regions. 
Recent high-resolution imaging at $\lambda = 
1.6~\micron$ by \citet{kudo08} discovered circumstellar nebulosity. 
The shape of the nebulosity is almost circular, and the azimuthally averaged radial profile of the 
brightness follows power-law forms of $r^{-2.5 \pm 0.1}$ in $110~\mathrm{AU} \le r \le 260~\mathrm{AU}$ and 
$r^{-6.5 \pm 0.2}$ in $280~\mathrm{AU} \le r \le 400~\mathrm{AU}$. 
Based on the fact that the exponent of the former law agrees with that observed in the nebulosity
around TW Hya \citep{wei02}, which can be explained in terms of scattering light from the surface of a 
flared disk \citep{kri00, tri01}, they argued that the origin of near-infrared nebulosity around \object{FN Tau} is also 
a flared circumstellar disk. 
Assuming that dust emission at 1.3 mm \citep[$31\pm9$ mJy; ][]{bec90} comes from the same 
regions as the nebulosity, the disk mass was estimated to be $7 \times 10^{-3}~M_{\sun}$.
To derive the disk parameters more directly, however, higher-resolution observations at submillimeter
wavelengths are essential. 

This paper presents simultaneous observations of dust continuum emission at 340 GHz and
CO ($J=3-2$) from \object{FN Tau} with the Submillimeter Array (SMA)\footnote{
The Submillimeter Array is a joint project between the Smithsonian Astrophysical Observatory 
and the Academia Sinica Institute of Astronomy and Astrophysics and is 
funded by the Smithsonian Institution and the Academia Sinica.}. 
High angular resolution provided by the SMA has allowed us to  
constrain better the distribution of circumstellar material.  
We describe the details of the observations in \S 2
and present the results in \S 3. 
Based on our results as well as those obtained by previous studies, 
we discuss in \S 4 the circumstellar structure of \object{FN Tau}, including the disk 
parameters and the origin of the near-infrared nebulosity. 

\section{Observations}

SMA observations of \object{FN Tau} were carried out on 2008 July 11 (UT). 
4K-cooled SIS mixer receivers were tuned at the central frequencies 336.00488 GHz 
and 345.53059 GHz for the lower and upper sidebands, respectively, and the continuum 
emission was received with a bandwidth of 1.7 GHz for each sideband. Data of CO ($J=3-2$) 
were simultaneously obtained with a frequency resolution of 101.5 kHz, corresponding 
to a velocity resolution of 0.088 km s$^{-1}$. The extended antenna configuration consisting 
of 7 antennas provided baselines ranging from 39 k$\lambda$ to 258 k$\lambda$. 
The largest brightness distribution to which these observations are sensitive enough 
(more than 50\% level) is $\sim 2.3\arcsec$ in FWHM when the 
distribution is Gaussian \citep{wil94}. Uranus was observed to 
obtain the passband characteristics of the system, while the two quasars 3C111 and 0336+323 
were observed every 15 minutes to track the time variation of the complex gain. 
The system temperature during the observations, $T_{\mathrm{sys}}$, was $230-450$ K. 

Visibility data were calibrated with the IDL MIR package, which 
was originally developed by the Owens Valley Radio Observatory as the MMA software 
package \citep{sco93} and later adapted for the SMA.  
After the correction for the visibility amplitude using $T_{\mathrm{sys}}$, the
gain table constructed from the 3C111 data was applied to the visibilities of \object{FN Tau} 
and 0336+323; the flux density of 3C111 at 340 GHz was derived to be 1.04 Jy from the 
comparison with Uranus. Calibrated visibilities of 0336+323 were carefully inspected to 
identify visibilities taken under poor conditions. All the visibilities identified 
were flagged out before further analyses in \S 3. Data from both the sidebands 
were combined in the analyses of the continuum emission, hence the effective central 
frequency is 340.767735 GHz. Imaging from the calibrated visibilities was made with 
Astronomical Image Processing System (AIPS) developed at the National Radio 
Astronomy Observatory (NRAO).

\section{Results}

\subsection{Continuum Emission}\label{cont-res}

Figure \ref{fig-fncont} shows the maps of the continuum emission from \object{FN Tau} 
constructed with the CLEAN algorithm under two different visibility weightings. Natural 
weighting provides a synthesized beam of $0\farcs78 \times 0\farcs71$ (P.A. = 38\degr) 
in FWHM, and the rms level in the emission-free regions is 1.7 mJy beam$^{-1}$, 
corresponding to 0.0324 K in brightness temperature. Gaussian fit to the detected emission 
reveals that the emission has a total flux density of $36.7 \pm 2.9$ mJy with no significant 
extension compared to the beam size. A sharper beam, $0\farcs74 \times 0\farcs57$ 
(P.A. = 68\degr) in FWHM, is given in the uniform weighting, but the detected emission 
is again point-like with a flux density of $31.5 \pm 4.7$ mJy. Although these results 
can be regarded as the detection of a point source, slightly smaller flux density with 
the uniform 
weighting may suggest that the spatial extent of the emission could marginally be 
resolved.  Similar comparison has also been made for 0336+323, which must be a perfect 
point source,  and the detected flux densities are $235.3 \pm 8.2$ mJy for the natural 
weighting and $233.8 \pm 7.8$  mJy for the uniform weighting; 
unlike the case of \object{FN Tau}, there is little difference between these two weightings. 

These situations can also be confirmed by the visibility plots (Figure \ref{fig-fnvis}). 
When a point source located at the phase center is observed, the real part of its visibility 
is a constant equal to its flux density while the imaginary part is zero for all the baselines. 
Indeed, these characteristics can be clearly seen in the visibility plots for 0336+323. 
Visibility plots for \object{FN Tau} also show similar characteristics, but the imaginary part 
at $\leq 165$ k$\lambda$ is nearly zero while the real part in $(135-165)$ k$\lambda$ is 
slightly smaller than that at shorter baselines. These can be explained by a circularly 
symmetric distribution of the emission with a finite extent. The submillimeter continuum 
emission from \object{FN Tau} is likely to have such a nature because the circumstellar 
nebulosity at near-infrared exhibits a circularly symmetric brightness distribution \citep{kudo08}. 
At baseline lengths greater than 180 k$\lambda$, however, the imaginary part has non-zero values, 
possibly due to higher phase noise. Thus the deviation from the perfect point-source case cannot 
confidently be identified in our results. 

Upper limit to the apparent size 
of the emission is estimated from a comparison with 
the analytic solution for the case of uniform brightness. Visibility amplitude for a disk with 
uniform brightness, $|V(D_{\lambda})|$, is expressed by
\begin{equation}
|V(D_{\lambda})| = F_0\left|\frac{J_1(2\pi D_{\lambda}\theta)}{\pi  D_{\lambda}\theta}\right|, 
\label{eqn:visamp}
\end{equation}
where $J_1$ is Bessel function of the first kind, $F_0$ and $\theta$ are the total flux density
and apparent radius of the disk, respectively, and $D_{\lambda}$ is projected baseline length 
in units of the observed wavelength \citep[e.g.,][]{wil94}. $|V(D_{\lambda})|$ peaks at $D_{\lambda} = 0$ with $F_0$ 
and takes the 80\% level of $F_0$ when $D_{\lambda}\theta \approx 0.21$. 
In the case of \object{FN Tau}, the real part of the visibility is $\sim 30$ mJy, or 80\% of 
the total flux (37 mJy), at $D_{\lambda} \sim 150$ k$\lambda$. This means $\theta =0.29\arcsec$, 
or 41 AU in radius, when Equation (\ref{eqn:visamp}) is applied. Indeed, the solution for 
$\theta =0.29\arcsec$ still traces the SMA data fairly well as shown in Figure \ref{fig-fnvis}(a). 
The outer radius of the detected emission is therefore estimated to be $\leq 41$ AU,  
significantly smaller than that of the near-infrared nebulosity \citep[$r= 260$ AU for 
the bright part;][]{kudo08}. 

Continuum flux density of \object{FN Tau} was also measured to be $31\pm 9$ mJy at 230 GHz 
with the $11\arcsec$ beam of the IRAM 30-meter telescope \citep{bec90}. Flux density measured 
at 340 GHz with the SMA (37 mJy)
is significantly smaller than the value simply extrapolated from the measurement 
at 230 GHz with $F_{\nu} \propto \nu^{2+\beta}$, where $\beta$ is the exponent of dust absorption 
coefficient for the disks around T Tauri stars 
ranging from 0 to 2 \citep[e.g., ][]{kita02, and07}. This suggests that 
some fraction of dust emission is missed in 
the SMA observations; this issue will be examined further in \S \ref{dustorigin}.

\subsection{CO Emission}

After averaging visibilities in five contiguous frequency channels, we made maps of 
CO ($J=3-2$) with a velocity resolution of 0.44 km s$^{-1}$. Natural weighting provides an 
rms noise level of 100 mJy beam$^{-1}$ with a synthesized beam of $0\farcs78 \times 0\farcs71$ 
(P.A. = 38\degr), or 2 K in brightness temperature, but no significant emission was detected. 

The disk component which is $\leq 41$ AU in radius and is 
detected with the dust continuum can also emit the radiation of CO 
($J=3-2$), but our sensitivity seems insufficient to reveal the CO emission as  
follows. When the disk is perfectly face-on and emits the
CO line whose brightness temperature is $\sim 19$ K, as suggested by the model fit 
in \S \ref{dustorigin}, beam dilution will make the observed intensity as low as 7.7 K, which is
just below the $4\sigma$ level. Note that this estimate is for an optimistic 
case because the emission from the disk is assumed to fall in a single velocity channel 
because of the perfect face-on configuration
\citep[e.g.,][]{omo92}. The Keplarian rotational velocity around \object{FN Tau} will be 
$\sim 1.5$ km s$^{-1}$ at $r=41$ AU, and the observed velocity width will be $\ga 1$  km s$^{-1}$
if the inclination angle is 20\degr, as inferred by \citet{kudo08}. 
In such a case, only a partial area will emit the CO radiation for a single velocity channel 
and the observed intensity will become lower. 
Furthermore, \object{FN Tau} is located in the L1495 cloud where the extended CO emission is 
very strong \citep[e.g.,][]{miz95} while the SMA is sensitive only to the increment from these 
extended components, which makes the detection even more difficult. 
More sensitive observations will be required to reveal the compact CO 
emission associated with the star. 

\section{Discussion}\label{discussion}

\subsection{Fitting of Compact Dust Emission with a Power-law Disk Model} \label{dustorigin}

\subsubsection{$\beta = 1$ case  \label{betaone}}

\begin{figure}
\epsscale{.80}
\plotone{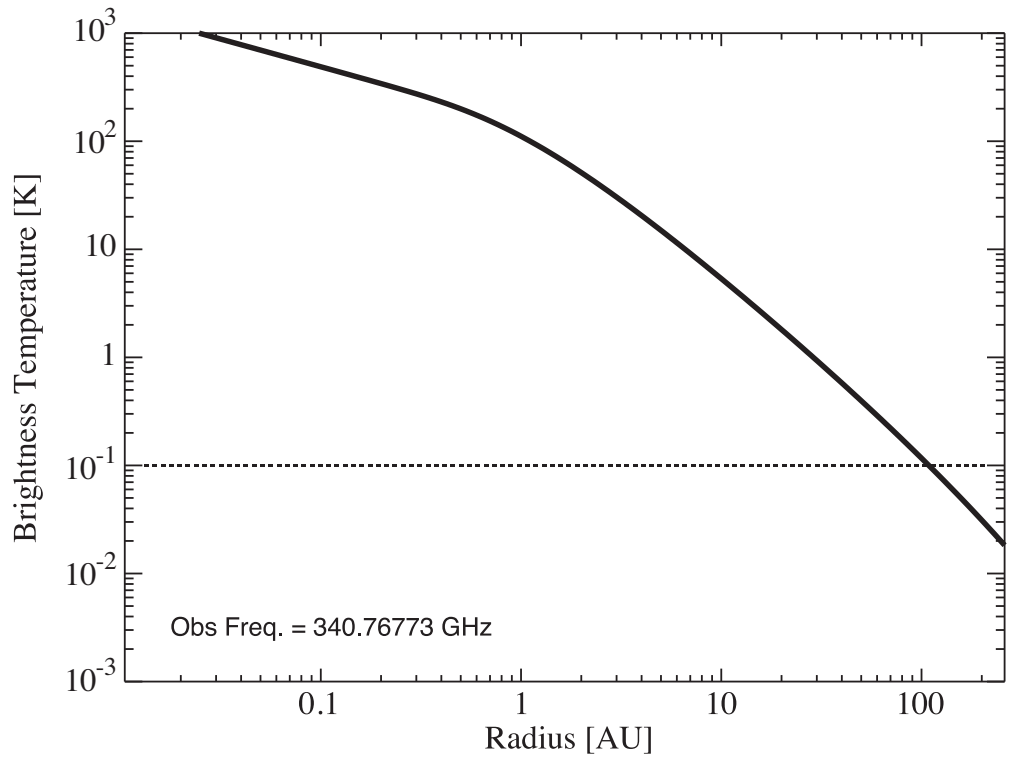}
\caption{Brightness distribution at $\nu = 340$ GHz of the model disk in 
\citet{kudo08}. Dotted line indicates the $3\sigma$ level of the continuum map 
with natural weighting, 0.097K (see \S \ref{cont-res}). \label{fig-fnbr}}
\end{figure}

The continuum emission detected with the SMA should originate from the disk around \object{FN Tau}.
\citet{kudo08} adopted a power-law disk model with inner and outer cutoffs \citep[e.g.,][]{kita02} and 
derived disk parameters by fitting the spectral energy distribution, including 
measurement at 230 GHz with the IRAM telescope \citep{bec90}. 
They assumed that the gas-to-dust mass ratio is 100
and that the mass opacity of dust particles, $\kappa _{\nu}$, 
at millimeter and submillimeter wavelengths can be expressed as 
\begin{equation}
 \kappa _{\nu} = 0.1 \left(\frac{\nu}{10^{12}~\mathrm{Hz}}\right)^{\beta}~\mathrm{cm}^{2}~\mathrm{g}^{-1}
\label{eqn:kappanu}
\end{equation}
with $\beta =1$. \citet{kudo08} also assumed 
that the bright inner part of the near-infrared nebulosity, $3.7\arcsec$ in diameter, 
coincides with the disk size.
The emission component detected with the SMA at 340 GHz, however, is much 
weaker than 100 mJy expected from the disk parameters obtained by \citet{kudo08}. 
Moreover, its size is so small that the $\sim 0.7\arcsec $ beam of the SMA could not 
resolve the spatial distribution. 
This is not due to an artifact caused by the finite sensitivity of our observations; 
as shown in Figure \ref{fig-fnbr},
the model disk derived by \citet{kudo08} has a brightness temperature at 340 GHz higher than 
the 3$\sigma$ level of our observations inside $\sim 100$ AU in radius,  
and the SMA would resolve the emission if the power-law disk had a radius greater than 100 AU. 
The present observations, therefore, prove that the assumption by \citet{kudo08} 
that the radius of the power-law disk coincides with the extent of near-infrared 
nebulosity is invalid.

\begin{figure}
\epsscale{1.0}
\plotone{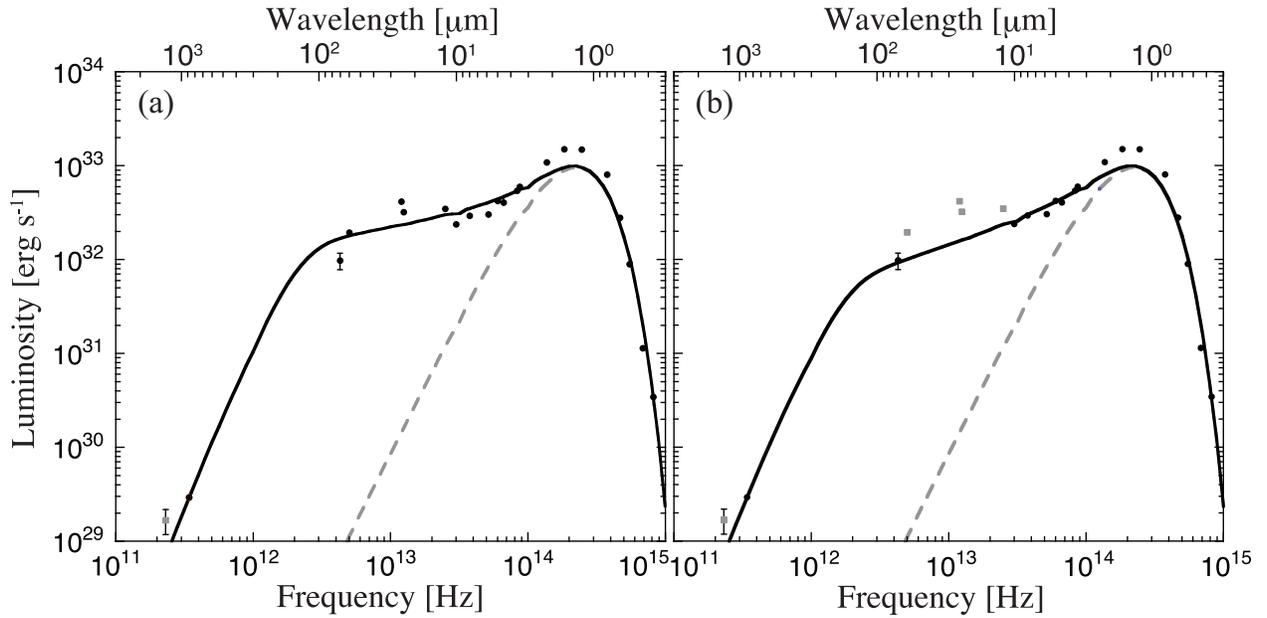}
\caption{Spectral energy distribution of FN Tau and that of a model disk which is 
best-fitted to
 ($a$) all the infrared data and ($b$) infrared data without $\lambda = (12-60)~\mu$m
under the assumption of $\beta=1$. 
The parameters of both the model disks are listed in Table \ref{tbl-1}, 
Gray dashed lines indicate the contribution of the stellar radiation. 
The data points indicated by gray squares are not used in the model fit. 
Error bars for data at $\lambda =70~\mu$m (20 \%) and at 1.3 mm (30 \%) are shown, 
but those for other data are not shown because each error is so small ($ <10$ \%) or 
is unknown. 
\label{fig-fnSED}}
\end{figure}

We therefore reexamine the disk parameters by fitting the spectral energy distribution 
with the same power-law disk model as that in \citet{kudo08} but 
under a different assumption that the outer radius of the disk is 41 AU, or the upper limit to 
the outer radius of the emission detected with the SMA (see \S \ref{cont-res}).
Indeed, the emission at $\lambda \le 60~\mu$m expected from 
the model disk in \citet{kudo08} is attributed to the region well inside 41 AU in radius. 
In the present model fit, we include the flux density at 340 GHz with the SMA, but
the result at 230 GHz obtained with the IRAM telescope \citep{bec90} is 
excluded because its beam size ($11\arcsec$ in FWHM) is so large that 
emission from a cold extended component outside the power-law disk may also be contained. 
At infrared wavelengths, 
photometric data obtained with the Spitzer Space telescope at 3.6, 4.5, 5.8, 8.0, 24 and 70 $\mu$m 
(Spitzer Taurus Catalog October 2008 v2.1)\footnote{The photometric data in 
the Spitzer Taurus Catalog are taken from http://irsa.ipac.caltech.edu/index.html.} 
are included in addition to those used by \citet{kudo08}, 
while the data point at 100 $\mu$m 
is excluded because this is merely an upper limit obtained with IRAS. 
The best fit disk parameters are summarized in the first row of Table \ref{tbl-1}, and the comparison between the 
spectral energy distribution of the model and the observed photometric data is shown in 
Figure \ref{fig-fnSED}($a$). Compared to the disk parameters inferred by \citet{kudo08}, 
the surface density distribution in Table \ref{tbl-1} is comparable while the temperature is slightly lower 
with a steeper radial distribution. Because of the smaller outer radius, 
the total disk mass is $1.1 \times 10^{-3}~M_{\sun}$, only 15\% of the estimate 
by \citet{kudo08}. 

Although the result shown in Figure \ref{fig-fnSED}($a$) is for the best fit case,
it still exhibits notable discrepancies in two wavelength ranges: 
(i) at near-infrared of $\lambda = 0.8-2.2~\mu$m, in which 
the photometric data are mainly taken from 
the 2MASS catalogue, and (ii) at wavelengths longer than $5.8~\mu$m. 
As shown in the following, the discrepancy at near-infrared is probably due to the 
contribution of the circumstellar nebulosity in scattering light, which is not included in the model. 
Photometric magnitude of FN Tau in the H-band from the 2MASS catalogue is 8.67 mag, corresponding 
to $6.4 \times 10^{-10}$ erg~s$^{-1}$~cm$^{-2}$ in $\lambda F_\lambda$, 
but this contains both the thermal radiation from the star-disk system and 
the scattering light from the circumstellar nebulosity. 
The contribution from the star in the model is calculated to be $3.7 \times 10^{-10}$ erg~s$^{-1}$~cm$^{-2}$, 
which is consistent with 
photometric measurements of M5 stars \citep{pic98}.
Thermal radiation from the model disk is estimated to be $\sim$ 10\% of the stellar radiation at this wavelength, 
hence the amount of discrepancy, or the contribution of scattering light, 
is derived to be $\sim 2.3 \times 10^{-10}$ erg~s$^{-1}$~cm$^{-2}$.
On the other hand, the flux density of the nebulosity in the H-band image by 
\cite{kudo08} is 6.9 mJy, or $1.3 \times 10^{-10}$ erg~s$^{-1}$~cm$^{-2}$. 
This is only $\sim$ 55 \% of the excess emission estimated above, but it
does not include the contribution from the area blocked by the occulting mask of the coronagraph 
($\sim 0.8\arcsec$ in diameter) and the spider pattern of the telescope. If the contribution from the 
blocked area is taken into consideration, the nebulosity component can account for the discrepancy at 
near-infrared in the model fit. 

The situation at longer wavelengths, which are important in deriving 
the disk temperature distribution, seems more complicated. 
Figure \ref{fig-fnSED}($a$) shows that the observed fluxes in $\lambda = 5.8-10~\mu$m are systematically 
smaller than the model while those in $\lambda = 12-60~\mu$m are  systematically 
larger than the model. These disagreements could be resolved if we adopt a more elaborate disk model 
with a flared surface layer of superheated dust grains \citep[e.g.,][]{chi97} because 
the surface layer can contribute additional fluxes mainly at the latter wavelength range. 
We should examine in detail these effects when we would 
investigate the vertical structure at smaller radii, but this is beyond the scope of this paper. 
To estimate the total mass or the surface density of the disk
based on our SMA results, 
a simple  power-law disk model may be acceptable because 
almost whole the disk emission at millimeter and submillimeter wavelengths 
are expected to originate from a single layer, i.e., 
the interior region near the mid-plane which is the dominant mass reservoir \citep{chi97}. 
For the temperature distribution of the interior region, 
a better estimate could be obtained with a model fit in which photometric
data that might be severely affected by the contribution from the surface layer
are eliminated. As shown in Figure \ref{fig-fnSED}($b$) and the fourth row of 
Table \ref{tbl-1}, the result of the model fit without data 
at $\lambda = 12-60~\mu$m, 
which exhibit significant upward deviations from the model  in Figure \ref{fig-fnSED}($a$), 
gives us a disk mass 40 \% higher than the case shown in Figure \ref{fig-fnSED}($a$)
due to the lower derived temperature. 
For the $\beta = 0$ and $\beta=2$ cases in \S \ref{otherbeta} and in Table \ref{tbl-1},
we will only show 
the results of the model fits in which all the infrared data are included, 
but the disk mass can be higher than those shown in 
Table \ref{tbl-1} by 40 \% in the model fits without data at $\lambda = 12-60~\mu$m, 
as similar to the $\beta = 1$ case. 

The disk models in Figures \ref{fig-fnSED}($a$) and  \ref{fig-fnSED}($b$) 
predict that the flux density 
at  230 GHz is $13 \pm 3$ mJy. This is significantly smaller than 
the flux density measured by the $11\arcsec$ 
beam of the IRAM telescope, $31\pm 9$ mJy \citep[][]{bec90}. Since
this measurement was made with a chopping position for the sky subtraction just $30\arcsec$ 
(4200 AU) apart from the star, contributions by the extended background should mostly be 
canceled out and the detected emission is likely to originate from 
some circumstellar components. One can also confirm easily that the contamination from the 
extended CO ($J=2-1$) line is negligible when one takes into account the fact that 
the frequency bandwidth of this observation was 50 GHz. 
The above comparison, therefore, suggests that the flux density measured with the IRAM telescope 
contains a slight excess of $18\pm 9$ mJy from the power-law disk 
model with $\beta = 1$, and that 
there is another component which is missed in the present SMA observations. 
The possible origin of this missing component, together with apparent disagreement 
in size between the emission seen in the SMA image and the near-infrared nebulosity, will 
be discussed in \S \ref{halo}. 

\subsubsection{Cases of Other $\beta$ Values  \label{otherbeta}}

\begin{figure}
\epsscale{.8}
\plotone{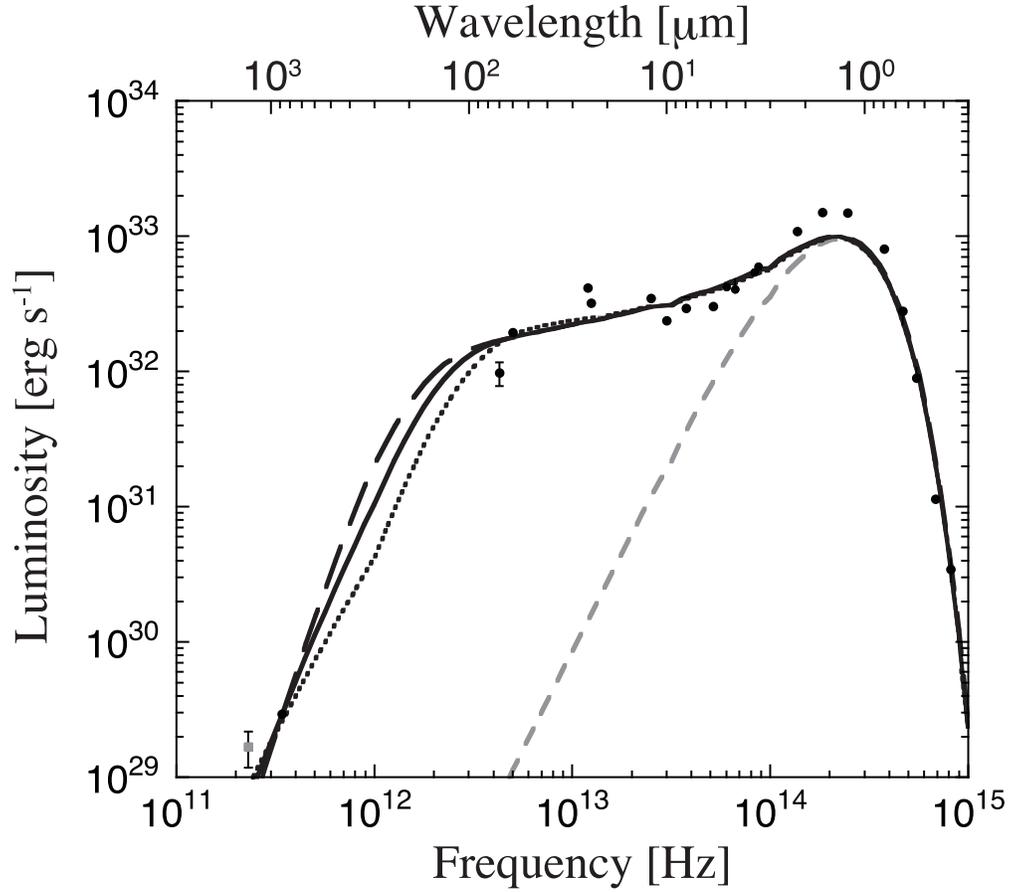}
\caption{Model fit of the spectral energy distribution of FN Tau by a
power-law disk model with different $\beta$ values; the $\beta=1$ case 
is shown by the solid line,  the $\beta=0$ case by the dotted line, and 
the $\beta=2$ case by the dashed line. 
The parameters of all the model disks are listed in Table \ref{tbl-1}. 
Gray dashed line indicates the contribution of the stellar radiation. 
The data point at $\lambda  = 1.3$ mm, indicated by a gray square, 
is not used in the model fit. 
\label{fig-SED-beta}}
\end{figure}

Since there is no photometric data at other millimeter and submillimeter wavelengths with a similar 
beam size as that of the SMA, we cannot directly determine $\beta$ inside the disk, or $r=41$ AU. 
We therefore made model fits with other $\beta$ values, 0 and 2, to evaluate uncertainties in 
derived physical quantities. The obtained disk parameters are 
summarized in Table \ref{tbl-1}, and the comparisons between the 
spectral energy distributions of the models and the observed photometric data are shown in 
Figure \ref{fig-SED-beta}. The parameters related to the temperature distribution in Table \ref{tbl-1} 
are almost insensitive to the change in $\beta$ because these are essentially determined by data points in the infrared 
range where the disk is optically thick. The surface density and disk mass, on the other hand, 
can be varied by a factor of 3 compared to the $\beta=1$ case according as the change of $\kappa_{\nu}$ 
following Equation (\ref{eqn:kappanu}). 

Although there are large uncertainties in the estimates for the disk mass and surface density, we 
can obtain several robust conclusions. First, the estimated disk mass is at most $5.9 \times 10^{-3}~ 
M_{\sun}$, which corresponds to 1.4 times the maximum mass in the $\beta = 2$ case 
of Table \ref{tbl-1}. This is just an order of Jupiter mass and seems 
insufficient to build a gas giant planet. As discussed by \citet{kudo08}, 
it is most likely that only small terrestrial or icy planets will form \citep{kokubo02}. 
Secondly, the disk size ($r\leq 41$ AU) is significantly smaller than the typical disk radius 
for solar-type T Tauri stars derived from a similar model fit including SMA data 
\citep[$r\sim 200$ AU; ][]{and07}, but the spectral energy distribution of FN Tau can be 
explained by a power-law disk model as similar to the cases of other solar-type T Tauri stars \citep{kita02,and07}. 
Together with the fact that the disk mass relative to the star ($0.3-3$ \%) 
is within the range of the statistics for solar-type T Tauri stars \citep{and07}, FN Tau has probably been formed 
through the standard process of star formation, i.e., the gravitational contraction of a cloud core, but not through 
dynamical ejection from a gravitationally unstable circumstellar disk \citep{bat02}.
Thirdly, all the best-fit model disks whose parameters are listed in Table \ref{tbl-1} cannot fully reproduce the 
flux density at $\lambda = 1.3$ mm obtained with the IRAM telescope 
\citep{bec90}, indicating that the two results with the SMA and the IRAM telescope 
cannot simultaneously be explained by a power-law disk only. 
This implies that there is another more extended component that cannot be well described 
by a power-law disk model and is missed in the SMA observations. 

\subsection{Nature of the Missing Extended Component \label{halo}} 

As shown in \S  \ref{dustorigin}, the best-fit power-law disk model cannot fully reproduce 
the flux density at 230 GHz obtained with the IRAM telescope in all the cases of $\beta = 0-2$. 
This implies that, in addition to the compact power-law component, there is a more extended 
component that is missed in the SMA observations; we will denote this component 
by ``halo'' in the following. 

\begin{figure}
\epsscale{.80}
\plotone{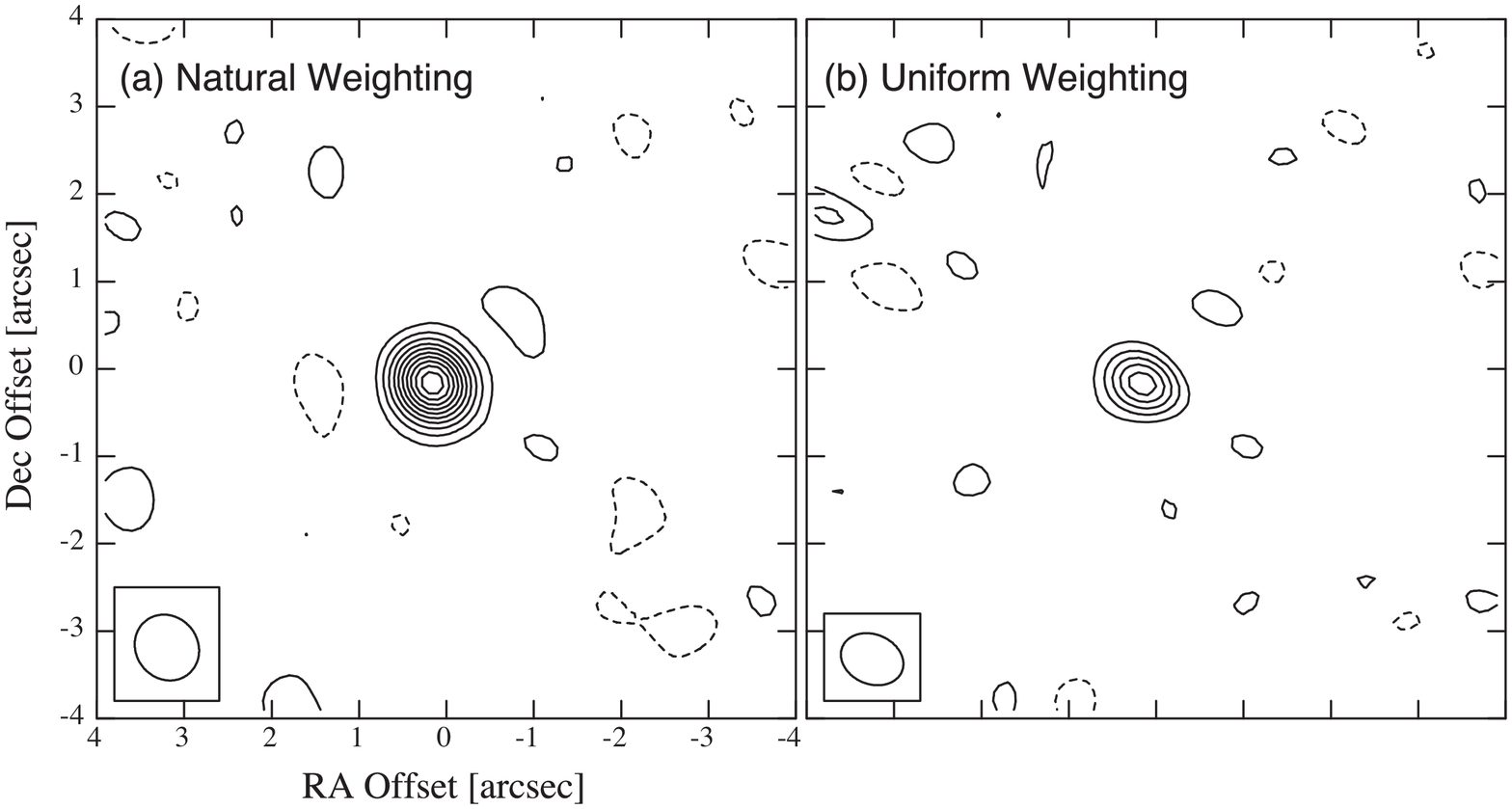}
\caption{Continuum emission at $\nu =340.768$ GHz toward 
FN Tau with the SMA. 
(a) Map with natural weighting. The synthesized beam is
$0\farcs78 \times 0\farcs71$ and P.A. = 38\degr ~in FWHM, and 
the rms noise level is 1.7 mJy beam$^{-1}$. 
(b) Map with uniform weighting. The synthesized beam is
$0\farcs74 \times 0\farcs57$ and P.A. = 68\degr ~in FWHM, and 
the rms noise level is 2.9 mJy beam$^{-1}$. 
Contour interval is $2\sigma$, 
starting at $ \pm 2\sigma$ levels. 
Dashed lines indicate negative levels. 
Ellipse at the bottom left corner of each panel shows the beam 
size in FWHM.
Both axes are measured from the phase center ($\alpha, \delta$) = 
(04$^\mathrm{h}$14$^\mathrm{m}$14\fs59, 28\degr27\arcmin58\arcsec) in J2000. 
\label{fig-fncont}}
\end{figure}

\begin{figure}
\epsscale{.80}
\plotone{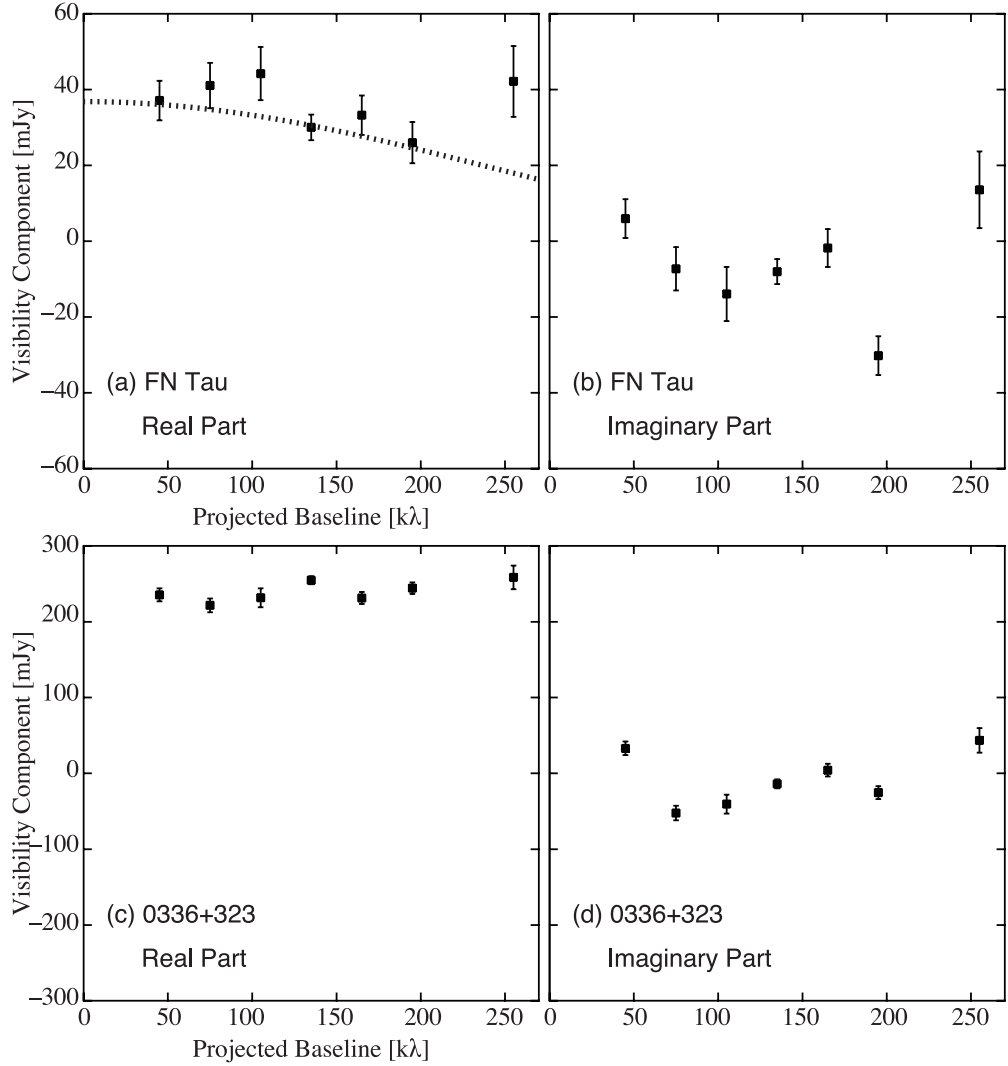}
\caption{Visibility plots for FN Tau (a, b) and 0336+323 (c, d). 
Both the real and imaginary parts of the visibilities are binned 
and averaged every 30 k$\lambda$. Standard errors are also plotted. 
For FN Tau, the positional offset of the emisson peak from 
the phase center, $(-0.2\arcsec, +0.2\arcsec)$, is corrected. 
No correction was made for 0336+323 because the emission peak 
coincides with the phase center. 
Dotted line in (a) indicates the analytic solution for the disk with uniform brightness 
whose total flux density is 37 mJy and the apparent radius is $0.29\arcsec$. 
\label{fig-fnvis}}
\end{figure}

We can roughly evaluate the lower limit to the size of the halo by requiring it not to be detected 
with our SMA observations. In the case of $\beta = 1$, for example, the flux density  
of the model disk component at 230 GHz is 13 mJy (see Table  \ref{tbl-1}), 
hence the median for the expected contribution of the halo at this frequency is 18 mJy.
Assuming that dust particles in the halo have the same $\beta$ as those in the disk, 
the flux density of the halo component at 340 GHz is estimated to be 58 mJy. 
If the emission of the halo is uniformly distributed inside $2.61\arcsec$ (i.e., 366 AU) in radius, 
it will have a brightness temperature equal to the $1\sigma$ level of the map in Figure \ref{fig-fncont}(a)
and cannot be detected. 
This estimate can also be verified in the visibility plot; 
the contribution of the halo at $D_{\lambda} = 45$ k$\lambda$ will be 3.5 mJy (i.e., 6\% of 
the total flux density of the halo) according to Equation (\ref{eqn:visamp}) with $\theta = 2.61\arcsec$, which is 
much smaller than the standard error of the visibility amplitude (see Figure \ref{fig-fnvis}).
We have summarized in Table \ref{tbl-halo} the lower limits to the size of the halo for 
other $\beta$ cases including their uncertainties. 

The estimated lower limits to the halo's radius in Table \ref{tbl-halo} are 
between 174 AU and 574 AU, depending on the assumed $\beta$ value. 
All of these numbers, however, are smaller than the beam size of the IRAM telescope, 
thus its photometric measurement can completely contain all the emission from the halo.
More importantly, the estimated size of the halo is quite reminiscent of the size of the near-infrared nebulosity: 
$1.85\arcsec$ (260 AU) in radius for the brighter part where the brightness distribution $S(r)$ follows 
$S(r) \propto r^{-2.5}$
and $2.85\arcsec$ (400 AU)  in radius for the tenuous periphery with $S(r) \propto r^{-6.5}$  \citep{kudo08}. 
These comparisons suggest that the circumstellar structure around \object{FN Tau} may consist of two 
components: a compact power-law disk ($r \leq 41$ AU) and a surrounding halo that probably corresponds to 
the origin of the near-infrared nebulosity. The SMA 
observations have revealed only the former component while both the components were detected 
by the IRAM telescope. 

When all the infrared data are used in a model fit,  
the best-fit parameters of the disk give $T \approx 19$ K at the outer edge of the disk, $r_{\mathrm{out}}= 41$ AU, 
regardless of the assumed $\beta$ (see Table \ref{tbl-1}). 
If we assume that the temperature in the halo, $T_{\mathrm{halo}}$, is also the same as
that at  $r_{\mathrm{out}}$ and uniformly 19 K, 
we can estimate its mass ($M_{\mathrm{halo}}$) and average surface density 
($\bar{\Sigma}_{\mathrm{halo}}$) by the following equations:
\begin{eqnarray}
M_{\mathrm{halo}} &=& \frac{F_{\nu}d^2}{\kappa_{\nu} B_{\nu}(T_{\mathrm{halo}})} \nonumber \\
&=& 3.8 \times 10^{-3} M_{\sun}\left(\frac{F_{340\mathrm{~GHz}}}{58~\mathrm{mJy}}\right)
\left(\frac{d}{140~\mathrm{pc}}\right)^2 
\left(\frac{\kappa _{\nu}}{0.0341~\mathrm{cm}^2 ~\mathrm{g}^{-1}}\right)^{-1}, \label{eqn-mhalo}\\
\bar{\Sigma}_{\mathrm{halo}} &=& \frac{M_{\mathrm{halo}}}{\pi r_{\mathrm{halo}}^2}
 = 7.9 \times 10^{-2} \mathrm{~g~cm}^{-2}
\left(\frac{M_{\mathrm{halo}}}{3.8 \times 10^{-3} M_{\sun}}\right)
\left(\frac{r_{\mathrm{halo}}}{366 \mathrm{~AU}}\right)^{-2}.\label{eqn-shalo}
\end{eqnarray}
$M_{\mathrm{halo}} $ and $\bar{\Sigma}_{\mathrm{halo}} $ for all the cases derived by 
Equation (\ref{eqn-mhalo}) or (\ref{eqn-shalo}) are shown in Table \ref{tbl-halo}, in which 
$\bar{\Sigma}_{\mathrm{halo}} $ are normalized by the disk surface density at  $r_{\mathrm{out}}$
(see Table \ref{tbl-1}). 
Note that $M_{\mathrm{halo}}$ in Table \ref{tbl-halo} will be higher by $\sim 40$ \% when 
we adopt the disk parameters obtained by a model fit without photometric data at $\lambda = 
12-60~\mu$m (see \S  \ref{betaone}), but the mass and surface density of the halo relative to the inner disk
remain unchanged because the disk mass and its surface density will also be higher by the same factor. 
Table \ref{tbl-halo} clearly indicates that the halo can contain mass comparable to 
or  at most 8 times greater than that of the inner power-law disk, 
but $\bar{\Sigma}_{\mathrm{halo}}$ should be lower 
than the surface density of the inner disk by an order of magnitude. 

Based on the fact that the surface brightness distribution of the near-infrared nebulosity, $S(r)$,
follows the relation $S(r) \propto r^{-2.5}$, 
\citet{kudo08} concluded that its origin is the scattering light at the surface of a flared disk. 
The new insight obtained with the present SMA observations, 
however, is that the circumstellar component outside at least 41 AU in radius cannot be 
described well in terms of a power-law disk model. 
Indeed, a power-law form of $S(r)$ can also be explained either by 
an optically thin, spherically symmetric envelope \citep[e.g.,][]{tam88, wei92} or 
by a flattened envelope with larger optical depth in almost pole-on configuration \citep{whi93}, 
though \citet{kudo08} did not examine these possibilities in detail. 
Among these possibilities, an optically thin and spherical envelope is unlikely; 
$\bar{\Sigma}_{\mathrm{halo}} =7.9 \times 10^{-2} \mathrm{~g~cm}^{-2}$ in 
Equation (\ref{eqn-shalo}) would correspond to 
$A_V \approx 9.4$ mag  to the star in the usual relation for the interstellar medium \citep{boh78}
when a spherical envelope is assumed, 
but this is much larger than the actual $A_V$ to the star \citep[1.35 mag; ][]{muz03}.

What is then the nature of the halo ? 
One possibility is that this is part of a flared circumstellar disk but 
the periphery regions that cannot be described well by a simple power-law disk model. 
Recently, \citet{hug08} investigated the apparent discrepancy between gas and dust 
outer radii derived from millimeter and submillimeter observations of protoplanetary disks.
They found that a disk model described by power laws in surface density and temperature 
that are truncated at an outer radius does not simultaneously reproduce the gas and 
dust emissions, but a simple accretion disk model that includes a tapered exponential 
edge in the surface density distribution does well. 
This is because the surface density in the outer tapered regions can be so low that the 
dust continuum emission at submillimeter is negligible but large enough to 
produce detectable emission in a probe of higher emission coefficient such as a rotational 
transition of CO or scattering light at near-infrared. In fact, 
one of the sample in \citet{hug08}, TW Hya, shows quite similar features to those seen in 
FN Tau; it exhibits near-infrared nebulosity whose size ($r=170$ AU)  
is significantly larger than the size of the submillimeter continuum emission 
determined by visibility fitting with a power-law disk model  ($r=60$ AU), 
and the radial brightness distribution of the nebulosity has a power-law form 
which is almost the same as that around FN Tau \citep{wei02}. 
The inferred surface density of the halo (see Equation (\ref{eqn-shalo}) and Table \ref{tbl-halo}), which is 
significantly lower than the inner power-law disk, is qualitatively similar to the outer regions 
of the model disk with an exponential tail. 
Another possibility is that the halo is a diffuse 
remnant of the protostellar envelope with flattened structure \citep{whi93}. 
The age of FN Tau is estimated to be $< 10^5$ yr 
\citep{muz03}, hence it is not surprising that FN Tau is still accompanied by such a component. 
Although these two candidates for the halo or the origin of the nebulosity are quite 
different in nature, 
it is unclear which is the case because both of them will have a scattering surface
that can account for the radial brightness profile of the near-infrared nebulosity.
Further sensitive observations in line emission will be critical to understand the nature of the halo.  

\acknowledgments

The authors would like to thank Misato Fukagawa for her comments on 
Spitzer observations that helped to improve the manuscript. 
We also thank an anonymous referee for providing invaluable suggestions.
This research was supported by Grants-in-Aid for
Scientific Research on Priority Areas, ``Development of Extra-Solar Planetary 
Science'', from the Ministry of Education, Culture, Sports, Science, and 
Technology (19015002), and by JSPS (18204017).
This work is also supported in part by a Grant-in-Aid for 
Scientific Research (A) from the Ministry of Education, 
Culture, Sports, Science and Technology of Japan (19204020).
NO is partially supported by the National Science Council (97-2112-M-001-019-MY2).
 
{\it Facilities:} \facility{SMA (CfA and ASIAA)}.

\clearpage



\clearpage

\clearpage

\clearpage

\clearpage

\clearpage

\begin{table}
\caption{Best Fit Parameters for a Model Disk \label{tbl-1}}
\begin{tabular}{c|ccccc}
\tableline\tableline
& $T_{\mathrm{1AU}}$ & $q$ & $\Sigma_{\mathrm{1AU}}$\tablenotemark{a} & $M_{\mathrm{disk}}\tablenotemark{a} $ 
& $F_{\mathrm{230GHz}}$  \\
& (K) &  & (g cm$^{-2}$) & ($M_{\sun}$)& (mJy) \\
\tableline
$\beta= 1\tablenotemark{b}$ &$141\pm3$ &$-0.543\pm0.011$ & $38\pm9$  &$(1.11\pm0.27)\times10^{-3}$  & $13\pm3$  \\
$\beta= 0\tablenotemark{b}$ &$144\pm3$ &$-0.534\pm0.012$ &$11\pm3$  &$(3.23\pm0.87)\times10^{-4}$  & $16\pm3$   \\
$\beta= 2\tablenotemark{b}$ &$141\pm3$ &$-0.545\pm0.011$ & $115\pm28$    &$(3.37\pm0.83)\times10^{-3}$  & $9\pm2$  \\
\tableline 
$\beta= 1\tablenotemark{c}$ &$126\pm5$ &$-0.577\pm0.015$ & $53\pm14$  &$(1.55\pm0.40)\times10^{-3}$  & $13\pm3$  \\
\tableline
\end{tabular}
\tablenotetext{a}{Gas-to-dust mass ratio is assumed to be 100. }
\tablenotetext{b}{Fitting results with all the infrared data.}
\tablenotetext{c}{Fitting results without using data at $\lambda = (12-60)~\mu$m.}
\tablecomments{All the uncertainties in derived physical quantities are evaluated so that 
each photometric measurement contains at most 20\% uncertainty, which is the case for 
the data point at $\lambda = 70 \micron$; Actual uncertainties for other data points 
at other wavelengths are much smaller than 20\%. 
Fixed parameters are as follows: $i = 20^{\circ}$, 
$r_{\mathrm{in}} = 0.015 \mathrm{AU}$, 
$r_{\mathrm{out}} = 41 \mathrm{AU}$, 
$r_{\mathrm{star}} = 0.0105 \mathrm{AU}$, 
$p=1$, 
$T_{\mathrm{star}} = 3240 \mathrm{K}$, and
$A_{V} = 1.35 \mathrm{mag}$. 
}
\end{table}

\clearpage

\begin{table}
\caption{Estimated Quantities for the Halo \label{tbl-halo}}
\begin{tabular}{c|ccccccc}
\tableline\tableline
& $F_{\mathrm{230GHz}}$ & $F_{\mathrm{340GHz}}$ & Minimum Radius & $M_{\mathrm{halo}}\tablenotemark{a}$ 
& $M_{\mathrm{halo}}/M_{\mathrm{disk}}$  & $\bar{\Sigma}_{\mathrm{halo}} /\Sigma_{\mathrm{disk}}$\tablenotemark{b} \\
& (mJy) &  (mJy)  & ($\arcsec$) & ($M_{\sun}$) & & \\
\tableline
$\beta= 1$ &$18\pm9$ &$58\pm29$ & $2.61^{+ 0.59}_{- 0.76}$ &$(3.8\pm1.9)\times10^{-3}$  & $3.4\pm1.7$ & $<0.086$ \\
\\
$\beta= 0$ &$15\pm9$ &$33\pm20$ & $1.96^{+ 0.52}_{- 0.72}$  &$(7.3\pm4.4)\times10^{-4}$  & $2.2\pm1.3$  &  $<0.10$ \\
\\
$\beta= 2$ &$21\pm9$ &$100\pm43$ &$3.43^{+ 0.67}_{- 0.84}$  &$(1.9\pm0.8)\times10^{-2}$  & $5.7\pm2.4$  &  $<0.083$ \\
\tableline
\end{tabular}
\tablenotetext{a}{Assuming that the temperature of the halo to be 19 K and that the gas-to-dust mass ratio to be 100.}
\tablenotetext{b}{Surface density at the outer radius (= 41 AU) is adopted as $\Sigma_{\mathrm{disk}}$.}
\end{table}


\begin{thebibliography}{}
\bibitem[Andrews \& Williams(2005)]{and05} Andrews, S. M., \& Williams, J. P. 2005, \apj, 631, 1134
\bibitem[Andrews \& Williams(2007)]{and07} Andrews, S. M., \& Williams, J. P. 2007, \apj, 659, 705
\bibitem[Apai et al.(2005)]{apai05}  Apai, D., Pascucci, I., Bouwman, J., Natta, A., Henning, Th., \& 
Dullemond, C.P. 2005, Science, 310, 834
\bibitem[Bate, Bonnell,  \& Bromm(2002)]{bat02} Bate, M. R., Bonnell, I. A., \& Bromm, V. 2002, \mnras, 332, L65 
\bibitem[Beaulieau et al.(2006)]{bea06} Beaulieau, J.~-P. et al. 2006, \nat, 439, 437
\bibitem[Beckwith et al.(1990)]{bec90} Beckwith, S. V. W., Sargent, A. I., Chini, R. S., \& G\"{u}sten, R. 1990, 
\aj, 99, 924
\bibitem[Bohlin, Savage, \& Drake(1978)]{boh78} Bohlin, R. C., Savage, B. D., \& Drake, J. F. 1978, \apj, 224, 132
\bibitem[Chiang \& Goldreigh(1997)]{chi97} Chiang, E. I., \& Goldreigh, P. 1997, \apj, 490, 368 
\bibitem[Forrest et al.(2004)]{for04} Forrest, W. J., et al. 2004, \apjs, 154, 443
\bibitem[Ho, Moran, \& Lo(2004)]{sma04} Ho, P. T. P., Moran, J. M., \& Lo, K. Y. 2004, \apj, 616, L1
\bibitem[Hughes et al.(2008)]{hug08} Hughes, A. M., Wilner, D. J., Qi, C. \& Hogerheijde, M. R. 2008, \apj, 678, 1119
\bibitem[Ida \& Lin(2004)]{ida04}Ida, S., \& Lin, D. N. C. 2004, \apj, 604, 388
\bibitem[Ida \& Lin(2005)]{ida05}Ida, S., \& Lin, D. N. C. 2005, \apj, 626, 1045
\bibitem[Johnson et al.(2007)]{joh07} Johnson, J. A., et al. 2007, \apj, 670, 833 
\bibitem[Kessler-Silacci et al.(2007)]{kes07} Kessler-Silacci, J. E. et al.  2007, \apj, 659, 680
\bibitem[Kitamura et al.(2002)]{kita02} Kitamura, Y., Momose, M., Yokogawa, S., Kawabe, R., Tamura, M., 
\& Ida, S. 2002, \apj, 581, 357
\bibitem[Kokubo \& Ida(2002)]{kokubo02} Kokubo, E., \& Ida, S. 2002, \apj, 581, 666
\bibitem[Krist et al.(2000)]{kri00} Krist, J. E., Stapelfeldt, K. R., Menard, F., Padgett, D. L., \& Burrows, C. J. 
2000, \apj, 538, 793 
\bibitem[Kudo et al.(2008)]{kudo08} Kudo, T. et al. 2008, \apjl, 673, L67
\bibitem[Laughlin et al.(2004)]{lau04} Laughlin, G., Bodenheimer, P., \&
Adams, F. C. 2004, \apjl, 612, L73 
\bibitem[Luhman et al.(2007)]{luh07} Luhman, K. L. et al., 2007, \apj, 666, 1219
\bibitem[Marcy et al.(2005)]{marcy05}Marcy, G., Butler, R. P., Fischer, D., Vogt, S., Wright, J. T., Tinney, C. G., 
\& Jones, H. R. A. 2005, Prog. Theor. Phys. Suppl., 158, 24 
\bibitem[Mayor et al.(2005))]{mayor05} Mayor, M., Pont, F., \& Vidal-Madjar, A. 2005, 
Prog. Theor. Phys. Suppl., 158, 43
\bibitem[Mizuno et al.(1995)]{miz95} Mizuno, A., Onishi, T., Yonekura, Y., Nagahama, T., Ogawa, H., 
\& Fukui, Y. 1995, \apjl, 445, L165
\bibitem[Mokler \& Stelze(2002)]{xray02} Mokler, F., \& Stelze, B. 2002, \aap, 391, 1025
\bibitem[Muzerolle et al.(2003)]{muz03}Muzerolle, J., Hillenbrand, L., Calvet, N., Briceno, C., \& Hartmann, L. 
2003, \apj, 592, 266
\bibitem[Natta et al.(2007)]{nat07} Natta, A., Testi, L., Calvet, N., Henning, Th., Waters, R., \& Wilner, D. 2007
 in Protostars and Planets V, ed. B. Reipurth, D. Jewitt, \& K. Keil (Tucson, AZ: 
University of Arizona Press), 767 
\bibitem[Omodaka et al.(1992)]{omo92} Omodaka, T., Kitamura, Y., \& Kawazoe, E. 1992, \apjl, 396, L87
\bibitem[Pascucci et al.(2009)]{pas09} Pascucci, I. et al. 2009, \apj, 696, 143
\bibitem[Pickles(1998)]{pic98} Pickles, A. J. 1998, \pasp, 110, 863
\bibitem[Sargent et al.(2006)]{sar06} Sargent, B. et al. 2006 \apj, 645, 395 
\bibitem[Scoville et al.(1993)]{sco93} 
Scoville, N. Z., Carlstrom, J. E., Chandler, C. J., Phillips, J. A., Scott, S. L., Tilanus, R. P. 
J., \& Wang, Z. 1993, \pasp, 105, 1482
\bibitem[Scholz et al.(2006)]{sch06} Scholz, A., Jayawardhana, R., \& Wood, K. 2006, \apj, 645, 1498
\bibitem[Strom et al.(1989)]{strom89}Strom, K. M., Strom, S. E., Suzan, E., Cabrit, S., \& Skrutskie, M. F. 1989, \aj, 97, 1451
\bibitem[Tamura et al.(1988)]{tam88} Tamura, M. et al. 1988, \apjl, 326, 17
\bibitem[Trilling et al.(2001)]{tri01} Trilling, D. E., Koerner, D. W., Barnes, J. W., Ftaclas, C., 
\& Brown, R. H. 2001, \apj, 552, L151
\bibitem[Weinberger et al.(2002)]{wei02} Weinberger, A. J., et al. 2002, \apj, 566, 409
\bibitem[Weintraub et al.(1992)]{wei92} Weintraub, D. A., Kastner, J. H., Zuckerman, B., 
\& Gatley, I. 1993, \apj, 391, 784
\bibitem[Whitney \& Hartmann(1993)]{whi93} Whitney, B. A., \& Hartmann, L. 1993, \apj, 402, 605 
\bibitem[Wilner \& Welch(1994)]{wil94} Wilner, D. J., \& Welch, W. J. 1994, \apj, 427, 898 
\end{thebibliography}
\end{document}